\documentclass[aps,prl,preprint,groupedaddress]{revtex4-1}

\usepackage[dvips]{graphicx} 
\begin{document}


\title{
Shape of Growth Rate Distribution \\
Determines the Type of Non-Gibrat's Property
}


\author{Atushi Ishikawa}
\email[ishikawa@kanazawa-gu.ac.jp]{}

\author{Shouji Fujimoto}
\email[fujimoto@kanazawa-gu.ac.jp]{}
\affiliation{Kanazawa Gakuin University, Kanazawa, Japan}

\author{Takayuki Mizuno}
\email[mizuno@ier.hit-u.ac.jp]{}
\affiliation{Hitotsubashi University, Tokyo, Japan}


\date{\today}

\begin{abstract}
In this study, the authors examine exhaustive business data on Japanese firms, 
which cover nearly all companies in the mid- and large-scale ranges in terms of firm size,
to reach several key findings on profits/sales distribution and business growth trends.
First, detailed balance is observed not only in profits data but also in sales data.
Furthermore, the growth-rate distribution of sales has wider tails than the linear
growth-rate distribution of profits in log-log scale.
On the one hand, in the mid-scale range of profits, the probability of positive growth 
decreases and the probability of negative growth increases symmetrically
as the initial value increases. This is called Non-Gibrat's First Property.
On the other hand, in the mid-scale range of sales, the probability of positive growth
decreases as the initial value increases, while the probability of negative growth 
hardly changes. This is called Non-Gibrat's Second Property.
Under detailed balance, 
Non-Gibrat's First and Second Properties are analytically derived from 
the linear and quadratic growth-rate distributions in log-log scale, respectively.
In both cases, the log-normal distribution is inferred from Non-Gibrat's Properties
and detailed balance.
These analytic results are verified by empirical data.
Consequently, this clarifies the notion that
the difference in shapes between growth-rate distributions of sales and profits
is closely related to 
the difference between the two Non-Gibrat's Properties in the mid-scale range.
\end{abstract}


\maketitle

\section{Introduction}
Distributions with a power-law tail have been found in various fields of 
natural and social science.
Examples of such studies include, for instance, avalanche sizes in a sandpile model \cite{Bak},
fluctuations in the intervals of heartbeats \cite{Peng},
fish school sizes \cite{Bonabeau}, 
citation numbers of scientific papers \cite{Render},
frequency of jams in Internet traffic \cite{TTS},
city sizes (see the recent review in Ref.~\cite{Saichev}), 
land prices \cite{Kaizoji}--\cite{Ishikawa100},
stock market price changes \cite{Mantegna},
and firm sizes \cite{Stanley00}.
Here, variables (denoted by $x$) follow the probability density function (PDF):
\begin{eqnarray}
    P(x) \propto x^{-(\mu+1)}~~~~{\rm for }~~~~x > x_{\rm th}~
    \label{Pareto}
\end{eqnarray}
over some size threshold $x_{\rm th}$.
This is called Pareto's Law, 
which was first observed in the field of personal income \cite{Pareto}.
The index $\mu$ is called the Pareto index.
Refer to Newman \cite{Newman} for a useful description of Pareto's Law.

In statistical physics,
the study of distributions with a power-law tail (\ref{Pareto}) is significant because
the $k$-th moment $\langle x^k \rangle = \int dx P(x) x^k$ diverges in the case of $\mu \le k$.
It is impossible to describe the system by using the variance $\sigma^2 = \langle x^2 \rangle$ 
or the standard deviation $\sigma$ in the case of $\mu \le 2$.
This feature comes from power-law behavior in the tail.
Furthermore,
it is worth noting that 
a large portion of the overall data are included in the power-law tail.
For example, approximately $90\%$ of total sales or profits in Japanese firms
are included in the power-law tail.
In economics (especially in macroeconomics), one of the major issues is
the state of the entire economy.
In this sense, it is important to clarify the nature of the power-law tail
not only in physics but also in economics.

In general, the power-law breaks below the size threshold $x_{\rm th}$
to suppress the divergence of the PDF \cite{Badger}, \cite{Montroll}.
There are many distributions that have a power-law tail.
These include, for instance, Classical Pareto Distribution (Pareto Type I Distribution),
Pareto Type II Distribution,
Inverse Gamma Distribution,
Inverse Weibull Distribution,
$q$--Distribution, A--Distribution and B--Distribution \cite{Aoyama book}.
In addition to these distributions, it has been hypothesized that many other distributions with a power-law tail follow the log-normal distribution for mid-sized variables below the size threshold $x_{\rm th}$: 
\begin{eqnarray}
    P(x) \propto \frac{1}{x} \exp \left[ - \frac{1}{2 \sigma^2} \ln^2 \frac{x}{\bar{x}} \right]~~~~{\rm for }~~~~x_{\rm min} < x < x_{\rm th}~.
    \label{Log-normal}
\end{eqnarray}
Here, $\bar{x}$ is a mean value and $\sigma^2$ is a variance.
A lower bound of the mid-scale range $x_{\rm min}$ is often related to the lower bound
of an exhaustive set of data.
A pseudo log-normal distribution is approximately derived from A--Distribution or B--Distribution
in the mid-sized range \cite{Aoyama book}.  

The study of distributions in the mid-scale range below the size threshold $x_{\rm th}$
is as important as the study of the power-law tail.
In physics, we are interested not only in the mechanism generating a power-law tail
but also in the reason for the tail breaking.
In economics, we should note that the majority of firms are mid-sized.
For instance, in sales or profits data, more than $90\%$ of the total number of firms 
are in the mid-scale range.
In this study, by examining exhaustive business data of Japanese firms 
that nearly cover the mid- and large-scale ranges,
the authors investigate the relevant distributions with a power-law tail.
This research is expected to be useful for understanding phenomena not only in economics
but also in physics.

On the one hand, 
it has been shown that Pareto's Law and the log-normal distribution can be derived
by assuming some model.
For example, a multiplicative process with boundary
constraints and additive noise can generate Pareto's Law \cite{Levy}.
On the other hand, by using no model, 
Fujiwara et al. have recently shown that Pareto's Law (\ref{Pareto}) is derived from 
Gibrat's Law and from the detailed balance observed in the large-scale range
of exhaustive business data \cite{Fujiwara}.
The relations among laws observed in exhaustive business data are important
for examining the characteristics of distributions based on firm-size.
For instance, in the study of Fujiwara et al.,
it was found that Pareto index $\mu$ is related to the difference between
a positive growth-rate distribution and a negative one.
Furthermore, along the lines of their study, 
one of the authors (A.~I) has shown that the log-normal distribution (\ref{Log-normal}) 
can be inferred from 
detailed balance and from Non-Gibrat's Property observed in the profits data of the mid-scale range 
\cite{Ishikawa}.
The study of the growth-rate distribution is an interesting subject in itself, and an ongoing investigation into this issue has progressed recently \cite{Riccaboni}.

Detailed balance means that the system is thermodynamically in equilibrium,
the state of which is described as
\begin{eqnarray}
    P_{J}(x_T, x_{T+1}) = P_{J}(x_{T+1}, x_T)~.
    \label{DetailedBalance}
\end{eqnarray}
Here, $x_T$ and $x_{T+1}$ are firm sizes at two successive points in time.
In Eq.~(\ref{DetailedBalance}), the joint PDF $P_{J}(x_T,x_{T+1})$ 
is symmetric under the time reversal exchange $x_T \leftrightarrow x_{T+1}$.

Gibrat's Law and Non-Gibrat's Property are observed in the distributions of 
firm-size growth rate $R=x_{T+1}/x_T$.
The conditional PDF of the growth rate $Q(R|x_T)$ is defined 
as $Q(R|x_T) = P_{J}(x_T,R)/P(x_T)$ by using the PDF $P(x_T)$ and 
the joint PDF $P_{J}(x_T,R)$.
Gibrat's Law, which is observed in the large-scale range,
implies that the conditional PDF $Q(R|x_T)$ is independent of the initial
value $x_T$ \cite{Gibrat}:
\begin{eqnarray}
    Q(R|x_T) = Q(R)~.
    \label{Gibrat}
\end{eqnarray}
Sutton \cite{Sutton} provides an instructive resource for obtaining the proper perspective on
Gibrat's Law.

Non-Gibrat's Property reflects the dependence of the growth-rate distribution
on the initial value $x_T$. 
The following properties are observed in the mid-scale range of positive profits data of 
Japanese firms \cite{Ishikawa}:
\begin{eqnarray}
    Q(R|x_T)&=&d(x_T)~R^{- t_{+}(x_T) - 1}~~~~~{\rm for}~~R > 1~,
    \label{FirstNon-Gibrat'sLaw1}
    \\
    Q(R|x_T)&=&d(x_T)~R^{+ t_{-}(x_T) - 1}~~~~~{\rm for}~~R < 1~,
    \label{FirstNon-Gibrat'sLaw2}
    \\
    t_{\pm}(x_T) &=& \pm \alpha~\ln x_T + C_{\pm}~.
    \label{FirstNon-Gibrat'sLaw3}    
\end{eqnarray}
Here, $\alpha$ and $C_{\pm}$ are positive constants.
In this composite Non-Gibrat's Property 
(\ref{FirstNon-Gibrat'sLaw1})--(\ref{FirstNon-Gibrat'sLaw3}), 
the probability
of positive growth decreases and the probability of negative growth increases  symmetrically
as the initial value $x_T$ increases in the mid-scale range.
It is particularly noteworthy that the shape of the growth-rate distribution 
(\ref{FirstNon-Gibrat'sLaw1})--(\ref{FirstNon-Gibrat'sLaw2}) uniquely determines
the change in the growth-rate distribution (\ref{FirstNon-Gibrat'sLaw3})
under detailed balance (\ref{DetailedBalance}).
Moreover, the rate-of-change parameter $\alpha$ appears in the log-normal distribution 
(\ref{Log-normal}).
We designate (\ref{FirstNon-Gibrat'sLaw1})--(\ref{FirstNon-Gibrat'sLaw3}) as 
Non-Gibrat's First Property 
to distinguish it from another Non-Gibrat's Property
that is observed in sales data.

The shape of the growth-rate distribution 
(\ref{FirstNon-Gibrat'sLaw1})--(\ref{FirstNon-Gibrat'sLaw2})
is linear in log-log scale.
This type of growth-rate 
distribution is observed in profits and income data of firms
(for instance \cite{Okuyama}, \cite{Ishikawa10}, \cite{Economics}).
In contrast, it has been reported in various articles that
the growth-rate distributions of assets, sales, number of employees in firms,
and personal income 
have wider tails than those of profits and income in log-log scale
(for instance \cite{Amaral}, \cite{Fujiwara}, \cite{Matia}, \cite{Fu}, 
\cite{Buldyrev}, \cite{Economics}).
In this case, the shape of the growth-rate distribution is different from 
Eqs.~(\ref{FirstNon-Gibrat'sLaw1}) and (\ref{FirstNon-Gibrat'sLaw2}).
There must be, therefore, another Non-Gibrat's Property corresponding to this shape.
In fact, it has been reported in several studies that
a Non-Gibrat's Property different from Non-Gibrat's First Property 
exists in the mid-scale range
of assets and sales of firms (for instance \cite{Aoyama}--\cite{Takayasu}).

In this study, we report the following findings by employing the sales data of Japanese firms, 
which include not only data in the large-scale range but also those in the mid-scale range.
\begin{enumerate}
  \item Detailed balance (\ref{DetailedBalance}) is confirmed in the mid- and large-scale ranges of sales data.
  \item In not only the large-scale range but also the mid-scale range of sales data, the growth-rate distributions have wider tails than those of profits in log-log scale.
  \item Under detailed balance (\ref{DetailedBalance}), the allowed change of the growth-rate distribution in the mid-scale range is analytically determined by using empirical data. The change is different from that of profits. We call this Non-Gibrat's Second Property.
  \item A log-normal distribution is derived from Non-Gibrat's Second Property and from detailed balance. This is verified with empirical data.
\end{enumerate}

From these results, we conclude that
the shape of the growth-rate distribution
determines the type of Non-Gibrat's Property in the mid-scale range.

\section{Non-Gibrat's First Property}
In this section, we review the analytic discussion in Ref.~\cite{Ishikawa}
and confirm it by applying the results to newly obtained data.
In the analytic discussion, detailed balance (\ref{DetailedBalance}) and 
the shape of the growth-rate distribution 
(\ref{FirstNon-Gibrat'sLaw1})--(\ref{FirstNon-Gibrat'sLaw2}) lead uniquely to 
a change in the growth-rate distribution (\ref{FirstNon-Gibrat'sLaw3}).
In addition, Non-Gibrat's First Property 
and detailed balance derive a log-normal distribution
(\ref{Log-normal}) in the mid-scale range.

In this study, we employ profits and sales data supplied by the Research Institute
of Economy, Trade and Industry, IAA (RIETI) \cite{RIETI}.
In this section we analyze profits data, and sales data are analyzed
in the next section.
The data set, which was created by TOKYO SHOKO
RESEARCH, LTD. \cite{TSR} in 2005, includes approximately
800,000 Japanese firms over a period of three years: the
current year, the preceding year, and the year before that.
The number of firms is approximately the same as the actual number of active Japanese firms.
This database is considered nearly comprehensive, at least in the mid- and
large-scale ranges.
In this study, we investigate the joint PDF $P_{J}(x_T,x_{T+1})$ and 
the distribution of the growth rate $R=x_{T+1}/x_T$.
Therefore, 
by using data of each firm in the previous three years,
we analyze 
a data set that has two values at two successive points in time as follows:
$(x_T, x_{T+1})$ 
= (data in preceding year, data in current year) $\cup$
(data in year before last, data in preceding year).
Here, $\cup$ indicates set-theoretic union.
This superposition of data is 
employed in order to secure a statistically sufficient sample size.
This procedure is
allowed in cases where
the economy is stable, that is, thermodynamically in equilibrium.
The validity is checked by detailed balance, as described below.
\begin{figure}[t]
\begin{center}
\includegraphics[height=6cm,width=6.5cm]{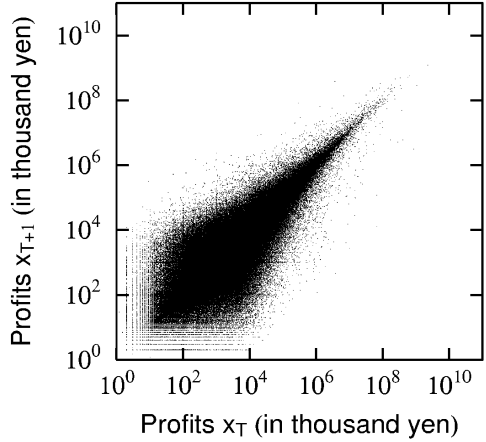}
\caption{\label{ProfitsDB-total} Scatter plot of positive profits in the database.
Here, $x_T$ and $x_{T+1}$ are positive profits of individual firms in consecutive years.}
\end{center}
\end{figure}

First, detailed balance (\ref{DetailedBalance}) is observed in profits data.
Note that only positive-profits data are analyzed here,
since we assume that non-negligible negative profits are not listed in the database.
Negative-profits data are thus not regarded as exhaustive.
We employ ``622,420'' data sets $(x_T, x_{T+1})$ that have two positive profits
at two successive points in time.
Figure.~\ref{ProfitsDB-total} shows the joint PDF $P_{J}(x_T,x_{T+1})$
as a scatter plot of individual firms.
Detailed balance (\ref{DetailedBalance}) is confirmed by the Kolmogorov--Smirnov (KS), Wilcoxon--Mann--Whitney (WMW), and Brunner--Munzel (BM) tests.
In the statistical tests, the range of $x_T$ is divided into $N$ bins
as $i_0 \le i_1 \le \cdots \le i_{n-1} \le i_{n} \le \cdots \le i_{N}$ 
to approximately equalize the number of data
in each bin ``$x_T \in [i_{n-1}, i_{n})$ and $x_T>x_{T+1}$.''
Here, $i_0$ and $i_N$ are the lower and the upper bounds of $x_T$, respectively.
We compare the distribution sample for 
``$P_{J}(x_T \in [i_{n-1}, i_{n}), x_{T+1})$ and $x_T>x_{T+1}$''
with another sample for 
``$P_{J}(x_T, x_{T+1} \in [i_{n-1}, i_{n}))$ and $x_T<x_{T+1}$''
($n = 1, 2, \cdots, N$) 
by making the null hypothesis that these two samples are taken from the same parent
distribution.

\begin{figure}[t]
\begin{center}
\includegraphics[height=6cm]{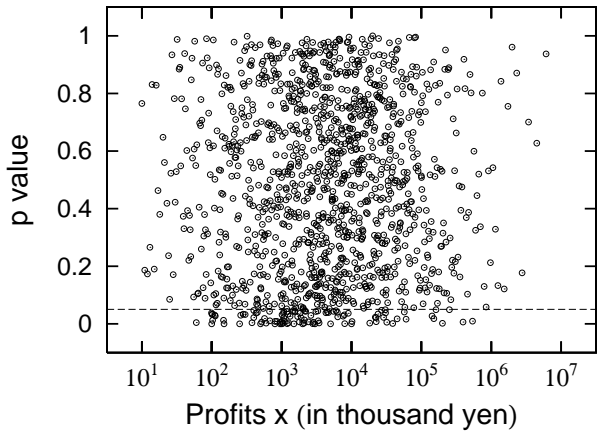}
\caption{\label{KStestProfits-total} Each $p$ value of the
WMW test for the scatter plot of positive-profits data points in Fig.~\ref{ProfitsDB-total}.}
\end{center}
\end{figure}

\begin{figure}[h!]
\begin{center}
\includegraphics[height=3cm]{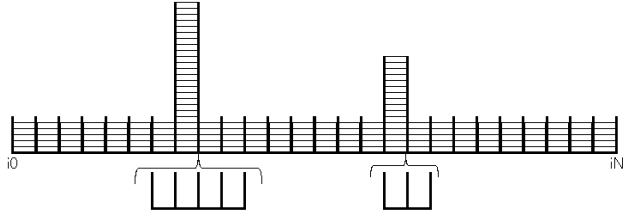}
\caption{\label{Fig} A bin with a round-number amount contains an exceptionally 
large number of data. 
In order to generally equalize the average amount of data in bins to the typical value, empty bins are inserted at bins with round-number amounts as needed.}
\end{center}
\end{figure}

Each $p$ value of the WMW test for the case of $N=2000$ is shown in Fig.~\ref{KStestProfits-total}.
Note that the profits data contain a large number of same-value amounts,
which are round numbers: $100$, $200$, $\cdots$, $1000$, $2000$, $\cdots$, 
$10000$, $20000$, $\cdots$.
This phenomenon is frequently observed in economic data.
A bin with a round-number amount may contain an exceptionally large number of data in this method of division.
For the case of $N = 2000$, almost all bins typically contain $200$ data; 
however, a bin with the round number of $5000$, for instance, 
contains an exceptional $4437$ data.
In order to generally equalize the average amount of data in bins to the typical value, 
an appropriate number of empty bins are inserted at such bins of round-number amounts as needed (Fig.~\ref{Fig}).
In the case of $N=2000$, there are $759$ empty bins.
$P$ values with respect to the remaining $1241$ bins are depicted 
in Fig.~\ref{KStestProfits-total},
in which $1141$ $p$ values exceed $0.05$.
Regardless of the division number $N$ and the kind of test, $p$ values exceed $0.05$ 
in approximately $92\%$ of bins.
This means that the null hypothesis is not rejected within the $5\%$ significance level
in approximately $92\%$ of the range.
This result does not change in the case where
the range of $x_T$ is divided into logarithmically equal bins.
Consequently, the detailed balance (\ref{DetailedBalance}) 
in Fig.~\ref{ProfitsDB-total} is generally confirmed.


\begin{figure}
\begin{center}
\includegraphics[height=6cm]{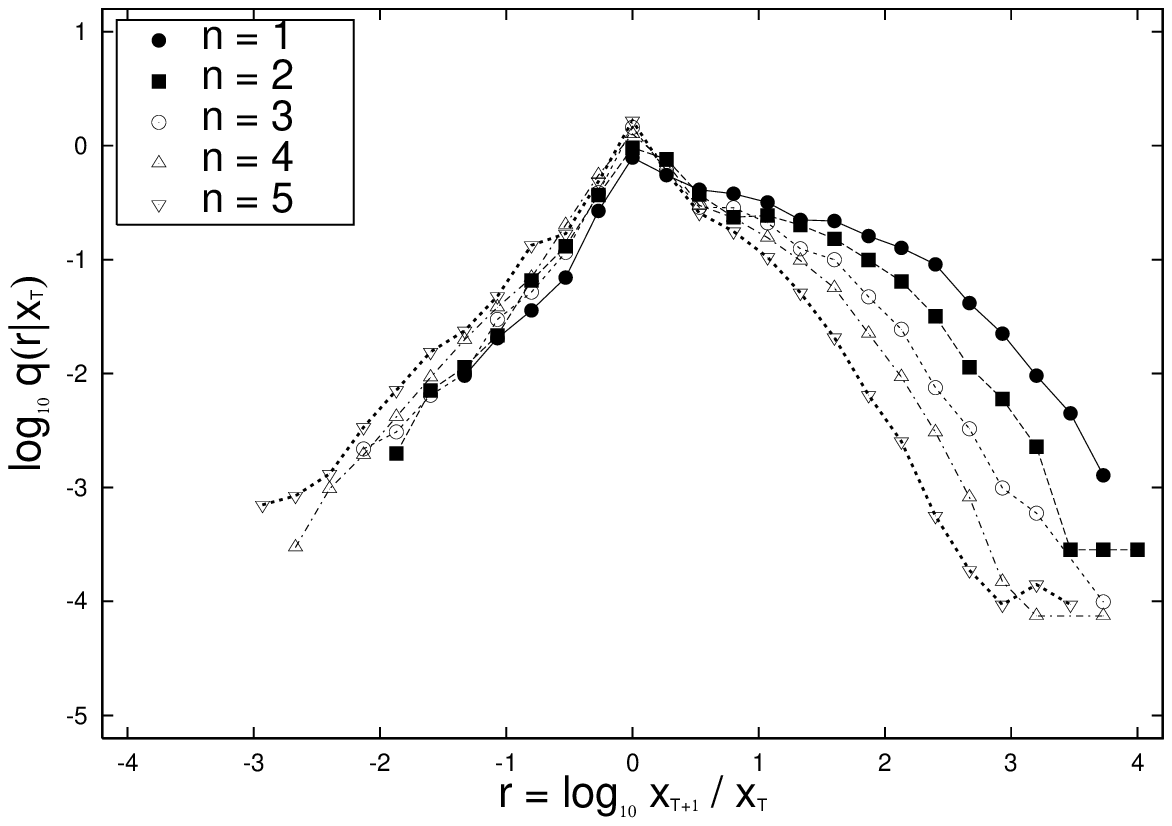}
\caption{\label{Profits-totalGrowthRate-1} Conditional PDFs of positive-profits growth rate 
in the low-scale range ($10^{1} \le x_T < 10^{3}$).
Here, $x_T$ and $x_{T+1}$ are positive profits in consecutive years, in thousand yen.}
\end{center}
\end{figure}

\begin{figure}
\begin{center}
\includegraphics[height=6cm]{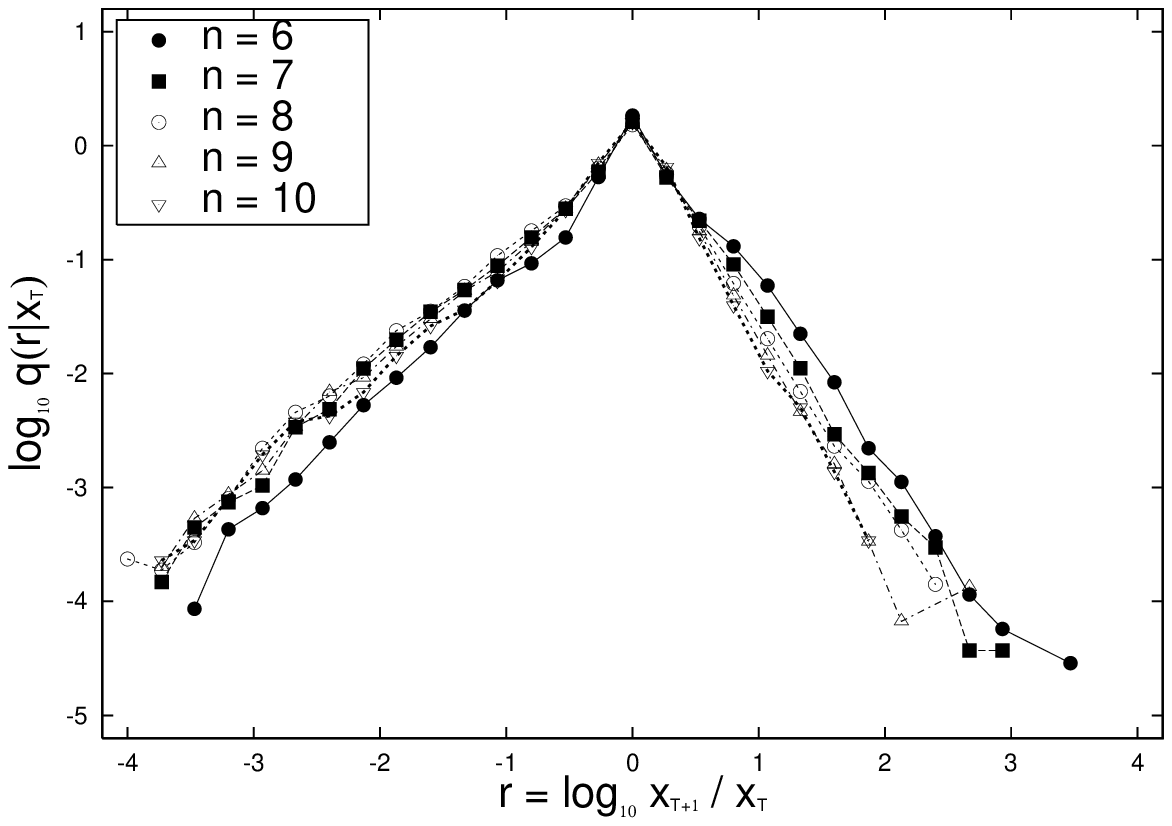}
\caption{\label{Profits-totalGrowthRate-2} Conditional PDFs of positive-profits growth rate 
in the mid-scale range ($10^{3} \leq x_T < 10^{5}$).
Here, $x_T$ and $x_{T+1}$ are positive profits in consecutive years, in thousand yen.}
\end{center}
\end{figure}

\begin{figure}
\begin{center}
\includegraphics[height=6cm]{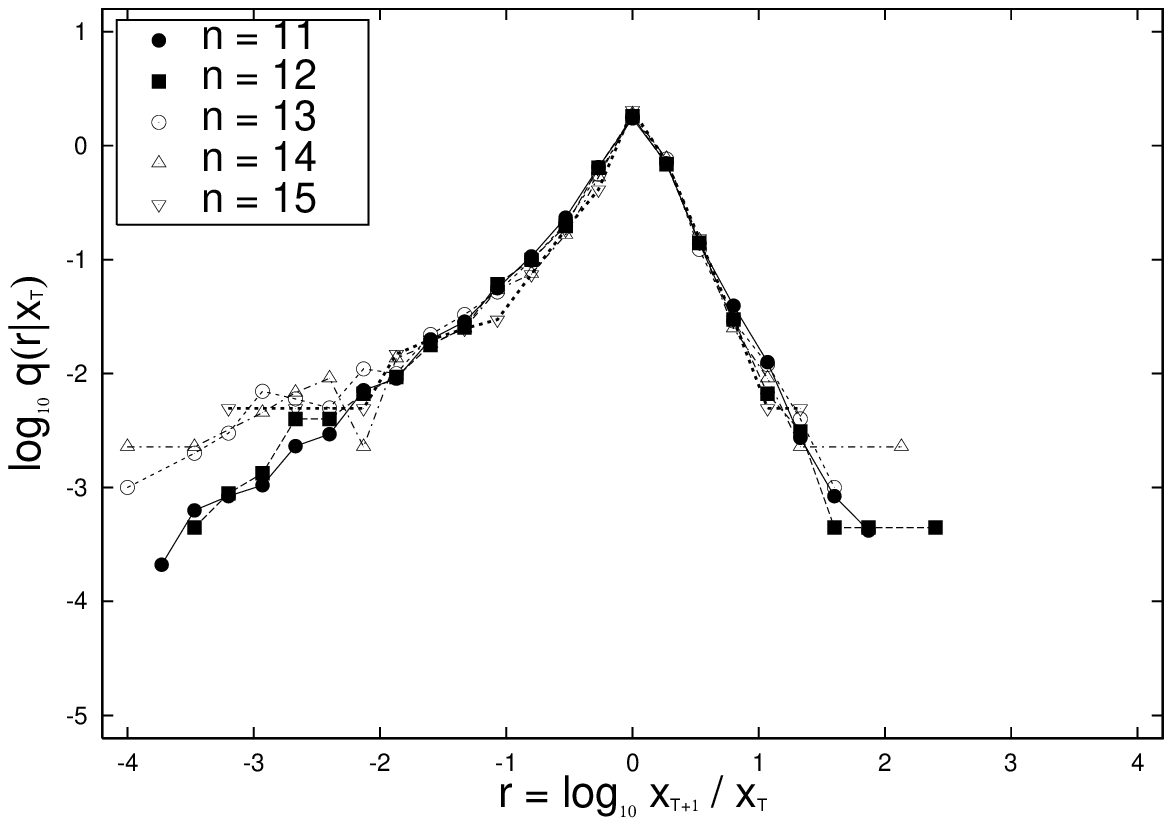}
\caption{\label{Profits-totalGrowthRate-3} Conditional PDFs of positive-profits growth rate 
in the large-scale range ($10^{5} \le x_T < 10^{7}$).
Here, $x_T$ and $x_{T+1}$ are positive profits in consecutive years, in thousand yen.}
\end{center}
\end{figure}

Second, we divide the range of the initial value $x_T$ into logarithmically equal bins as
$x_T \in [10^{1+0.4(n-1)},10^{1+0.4n})$ $(n=1,2,\cdots,15)$
in order to identify the shape of the growth-rate distribution and 
the change as the initial value $x_T$ increases.
The conditional PDFs $q(r|x_T)$ of the logarithmic growth rate
$r=\log_{10} R$ are shown in 
Figs.~\ref{Profits-totalGrowthRate-1}--\ref{Profits-totalGrowthRate-3}.
In Figs.~\ref{Profits-totalGrowthRate-2} and \ref{Profits-totalGrowthRate-3},
the growth-rate distributions in the mid- and large-scale ranges are
approximated by a linear function of $r$:
\begin{eqnarray}
    \log_{10}q(r|x_T)&=&c(x_T) - t_{+}(x_T)~r~~~~~{\rm for}~~r > 0~,
    \label{approximation1}
    \\
    \log_{10}q(r|x_T)&=&c(x_T) + t_{-}(x_T)~r~~~~~{\rm for}~~r < 0~.
    \label{approximation2}
\end{eqnarray}
The approximation (\ref{approximation1})--(\ref{approximation2}) is equivalent to
Eqs.~(\ref{FirstNon-Gibrat'sLaw1}) and (\ref{FirstNon-Gibrat'sLaw2})
by using relations $\log_{10} q(r|x_T) = \log_{10} Q(R|x_T) + r + \log_{10} (\ln 10)$ and
$d(x_T) = 10^{c(x_T)}/\ln10$. 
From $\int^{\infty}_{0} dR~Q(R|x_T)=1$, the normalization coefficient $d(x_T)$ 
(or the intercept $c(x_T)$) is determined as 
\begin{eqnarray}
    \frac{1}{d(x)} = \frac{1}{t_{+}(x)} + \frac{1}{t_{-}(x)}~.
    \label{dandt}
\end{eqnarray}

Following the discussion in a previous work~\cite{Ishikawa},
we derive the change in the growth-rate distribution (\ref{FirstNon-Gibrat'sLaw3})
from the shape of the growth-rate distribution 
(\ref{FirstNon-Gibrat'sLaw1})--(\ref{FirstNon-Gibrat'sLaw2})
under detailed balance (\ref{DetailedBalance}) and then
derive the log-normal distribution in the mid-scale range.
%
Under the exchange of variables from $(x_T, x_{T+1})$ to $(x_T,R)$, 
two joint PDFs $P_{J}(x_T, x_{T+1})$ and $P_{J}(x_T, R)$ are related to each other
as $P_{J}(x_T, R) = x_T P_{J} (x_T, x_{T+1})$.
Substituting the joint PDF $P_{J}(x_T, R)$ for the conditional PDF $Q(R|x_T)$
and using detailed balance (\ref{DetailedBalance}), we obtain
\begin{eqnarray}
    \frac{P(x_T)}{P(x_{T+1})} = \frac{1}{R} \frac{Q(R^{-1}|x_{T+1})}{Q(R|x_T)}~.
    \label{DetailedBalance2}
\end{eqnarray}
By substituting the conditional PDF for the shape of the growth-rate
distribution
(\ref{FirstNon-Gibrat'sLaw1})--(\ref{FirstNon-Gibrat'sLaw2}), 
another expression of detailed balance (\ref{DetailedBalance2}) is reduced to 
\begin{eqnarray}
    \frac{\tilde{P}(x_T)}{\tilde{P}(x_{T+1})} = R^{+t_{+}(x_T)-t_{-}(x_{T+1})+1}~
    \label{DetailedBalance3}
\end{eqnarray}
for the case of $R>1$.
Here, we denote $\tilde{P}(x) = d(x)~P(x)$.
By expanding Eq.~(\ref{DetailedBalance3}) around $R=1$ with $x_T \to x$ and $x_{T+1} \to R~x$, 
the following three differential equations are obtained:
\begin{eqnarray}
    \Bigl[1+t_{+}(x)-t_{-}(x) \Bigr] \tilde{P}(x) 
        + x~ {\tilde{P}}^{'}(x) = 0~,
    \label{DE1}
\end{eqnarray}
\begin{eqnarray}
    {t_{+}}^{'}(x)+{t_{-}}^{'}(x)=0~,~~~
    {t_{+}}^{'}(x)+x~{t_{+}}^{''}(x)=0~.   
    \label{DE2}
\end{eqnarray}
The same differential equations are obtained for $R < 1$.
Equations~(\ref{DE2}) uniquely fix $t_{\pm}(x_T)$ 
as Eq.~(\ref{FirstNon-Gibrat'sLaw3}).
Now, let us verify this by empirical data.

\begin{figure}[b]
\begin{center}
\includegraphics[height=6cm]{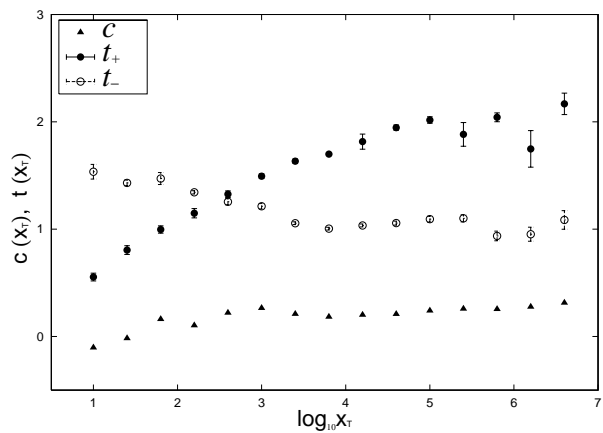}
\caption{\label{Profits-total-Evaluation} Estimations of $c(x_T)$ and $t_{\pm}(x_T)$.
Here, $x_T$ is the lower bound of each bin, in thousand yen, and
$c(x_T)$ is the original value of the growth-rate distribution.
From left, each point on the graph represents $n=1,2, \cdots, 15$.}
\end{center}
\end{figure}

Figure~\ref{Profits-total-Evaluation} shows $t_{\pm}(x_T)$ and $c(x_T)$
estimated by fitting the approximation (\ref{approximation1})--(\ref{approximation2})
to each growth-rate distribution in 
Figs.~\ref{Profits-totalGrowthRate-1}--\ref{Profits-totalGrowthRate-3}.
In Fig.~\ref{Profits-total-Evaluation}, $c(x_T)$ is fixed as the empirical value
and $t_{\pm}(x_T)$ is estimated by using the least-squares method.
In Fig.~\ref{Profits-totalGrowthRate-1}, 
the linear function (\ref{approximation1})--(\ref{approximation2})
is difficult to approximate for each growth-rate distribution,
and the values for $n=1,2,\cdots,5$ in Fig.~\ref{Profits-total-Evaluation} are untrustworthy.
In Fig.~\ref{Profits-totalGrowthRate-2}, however, the linear approximation 
(\ref{approximation1})--(\ref{approximation2}) is appropriate.
Applying the change in the growth-rate distribution 
$t_{\pm}(x_T)$ (\ref{FirstNon-Gibrat'sLaw3})
to $n=6, 7, 8$ $(10^{3} \le x_T < 10^{4.2})$ in Fig.~\ref{Profits-total-Evaluation},
we obtain the rate-of-change parameter $\alpha=0.11 \pm 0.02$ from $t_+(x_T)$
and $\alpha=0.11 \pm 0.03$ from $t_-(x_T)$ by using the least-squares method.
This coincidence of two estimated values guarantees Non-Gibrat's First Property 
(\ref{FirstNon-Gibrat'sLaw1})--(\ref{FirstNon-Gibrat'sLaw3}) in the empirical data.
We regard $10^{3} \le x_T < 10^{4.2}$ as the mid-scale range.

\begin{figure}[t]
\begin{center}
\includegraphics[height=6cm]{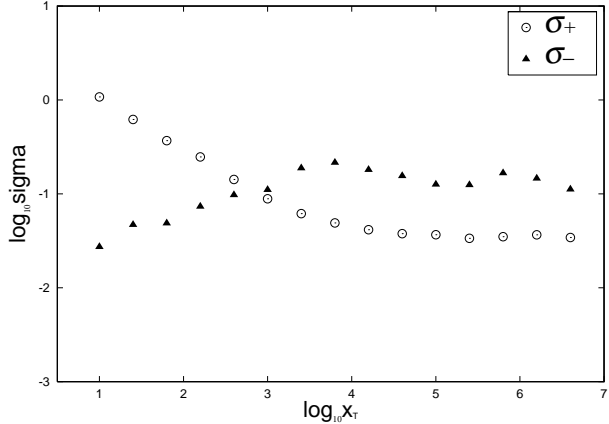}
\caption{\label{vvar} Estimations of $\sigma_{\pm}(x_T)$.
Here, $x_T$ is the lower bound of each bin, in thousand yen.
From left, each point on the graph represents $n=1,2, \cdots, 15$.}
\end{center}
\end{figure}

In Fig.~\ref{Profits-totalGrowthRate-3}, the growth-rate distribution barely changes
as $n$ increases.
This means that Gibrat's Law (\ref{Gibrat}) is valid in the large-scale range.
In Fig.~\ref{Profits-total-Evaluation}, the values $t_{\pm}(x_T)$ vary 
in the large-scale range,
since the number of data in Fig.~\ref{Profits-totalGrowthRate-3} 
is statistically insufficient to estimate $t_{\pm}(x_T)$
by the least-squares method.
However, by measuring the positive and negative standard deviations $\sigma_{\pm}$ of
each growth-rate distribution 
in Figs.~\ref{Profits-totalGrowthRate-1}--\ref{Profits-totalGrowthRate-3},
we confirmed that the growth-rate distribution only slightly changes in the range 
$x_T \ge 10^5$ (Fig.~\ref{vvar}). 
From Fig.~\ref{vvar}, we regard $x_T \ge 10^5$ as the large-scale range
and set $\alpha=0$ in this range.
Strictly speaking, a constant parameter $\alpha$ must not take different values.
However, in the database, a large number of firms stay in the same range
for two successive years.
This parameterization is, therefore, generally suitable for describing the PDF.

In Fig.~\ref{Profits-total-Evaluation},
$c(x_T) = \log_{10} (d(x_T)\ln10)$ hardly changes in the mid- and large-scale ranges $x_T \ge 10^3$.
This is consistent with $C_{\pm}>>\alpha \ln x_T$ 
in Eqs.~(\ref{FirstNon-Gibrat'sLaw3}) and (\ref{dandt}).
Consequently, by approximation we determine that the dependence of $d(x_T)$ on $x_T$ is negligible 
in the mid- and large-scale ranges.
Using $t_{\pm}(x)$ (\ref{FirstNon-Gibrat'sLaw3}), 
Eq.~(\ref{DE1}) uniquely decides the PDF of $x$ as
\begin{eqnarray}
    P(x) = C~{x}^{-(\mu+1)}
    ~\exp \left[ - \alpha \ln^2 x \right]
    ~~~~~{\rm for}~~x > x_{\rm min}~.
    \label{HandM}
\end{eqnarray}
Here, we regard $d(x)$ in $\tilde{P}(x)=d(x)~P(x)$ as a constant and 
denote $\mu=C_+ - C_-$.
The solutions (\ref{FirstNon-Gibrat'sLaw3}) and (\ref{HandM}) 
satisfy Eq.~(\ref{DetailedBalance3}) beyond perturbation around $R = 1$, and thus
these are not only necessary but also sufficient.

\begin{figure}[!h]
\begin{center}
\includegraphics[height=6cm]{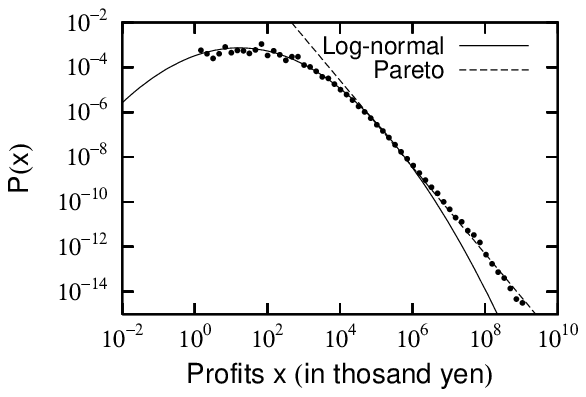}
\caption{\label{ProfitsDistribution-total} 
A PDF of positive profits in the database.
Pareto's Law is observed in the large-scale range ($x \ge 10^5$)
and in the log-normal distribution in the mid-scale range ($10^3 \le x < 10^{4.2}$).}
\end{center}
\end{figure}

Figure~\ref{ProfitsDistribution-total} shows that the resultant PDF (\ref{HandM})
fits correctly with the empirical profits data.
In the large-scale range ($\alpha=0$), 
the PDF (\ref{HandM}) behaves as Pareto's Law (\ref{Pareto}).
The Pareto index is estimated as approximately $\mu \sim 1$
in the large-scale range ($x \ge 10^5$) of Fig.~\ref{ProfitsDistribution-total}.
In the mid-scale range, the PDF (\ref{HandM}) behaves as the 
log-normal distribution (\ref{Log-normal})
with $\alpha = 1/(2 \sigma^2)$, $\mu = - \ln \bar{x}/(\sigma^2)$.
Applying the PDF (\ref{HandM}) to the mid-scale range ($10^{3} \le x < 10^{4.2}$)
of Fig.~\ref{ProfitsDistribution-total}, we obtain the rate-of-change parameter
$\alpha = 0.082 \pm 0.089$
by using the least-squares method.
The error bar is not small because we have applied the least-squares method 
to the quadratic curve in log-log scale.
The estimated value ($\alpha = 0.082 \pm 0.089$) is, however, consistent with 
the values estimated by the change in $t_{\pm}(x_T)$ ($\alpha = 0.11 \pm 0.02$ or 
$0.11 \pm 0.03$).
From these results, we conclude that Non-Gibrat's First Property is confirmed 
by the empirical data.
\section{Non-Gibrat's Second Property}

\begin{figure}
\begin{center}
\includegraphics[height=6cm]{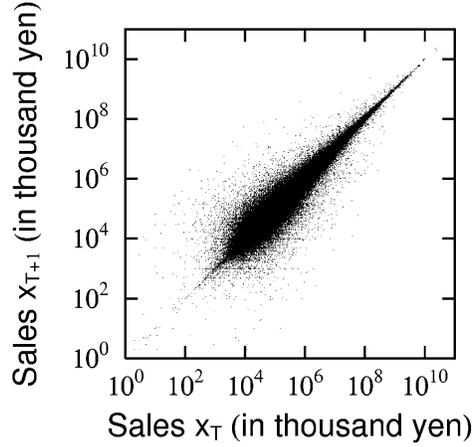}
\caption{\label{SalesDB-total} Scatter plot of sales in the database.
Here, $x_T$ and $x_{T+1}$ are sales of individual firms in consecutive years.}
\end{center}
\end{figure}

\begin{figure}
\begin{center}
\includegraphics[height=6cm]{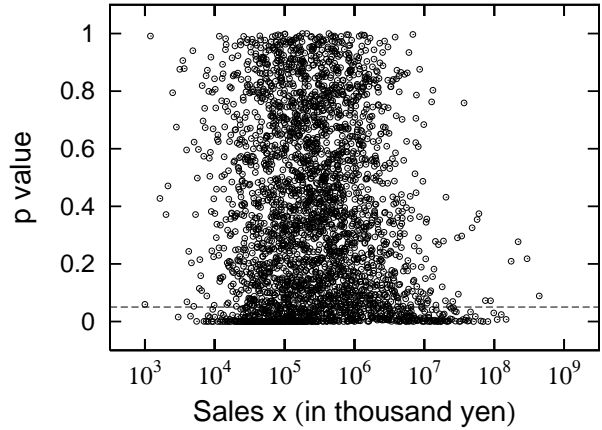}
\caption{\label{KStestSales-total} Each $p$ value of the BM test for the scatter plot of sales  data points in Fig.~\ref{SalesDB-total}.}
\end{center}
\end{figure}

In this section, we investigate another Non--Gibrat's Property 
observed in the mid-scale
range of sales data. This is the main aim of this study.
First, detailed balance (\ref{DetailedBalance}) is also observed in sales data.
Here, we employ ``1,505,108'' data sets $(x_T, x_{T+1})$ that have two sales
at two successive points in time.
Figure~\ref{SalesDB-total} shows the joint PDF $P_{J}(x_T,x_{T+1})$
as a scatter plot of individual firms.
Detailed balance (\ref{DetailedBalance}) is also confirmed by using the 
KS,
WMW, and BM tests
in the same manner as in the previous section.
Figure~\ref{KStestSales-total} shows each $p$ value of the BM test for the 
$N=5000$ case. 
Regardless of the division number $N$ and the kind of test, $p$ values exceed $0.05$ 
in approximately $82\%$ of bins.
This means that the null hypothesis is not rejected within the $5\%$ significance level
in approximately $82\%$ of the range.
Note that the sales data also contain a large number of same-value amounts, which are round numbers.
$P$ values of the statistical test for bins with a large number of round values
are unusually small.
In this situation, $82\%$ is acceptable.
The percentage is slightly higher in the case where the range of $x_T$ is divided into 
logarithmically equal bins.
We assume, therefore, that detailed balance (\ref{DetailedBalance}) 
in Fig.~\ref{SalesDB-total} is generally verified.

\begin{figure}[h]
\begin{center}
\includegraphics[height=6cm]{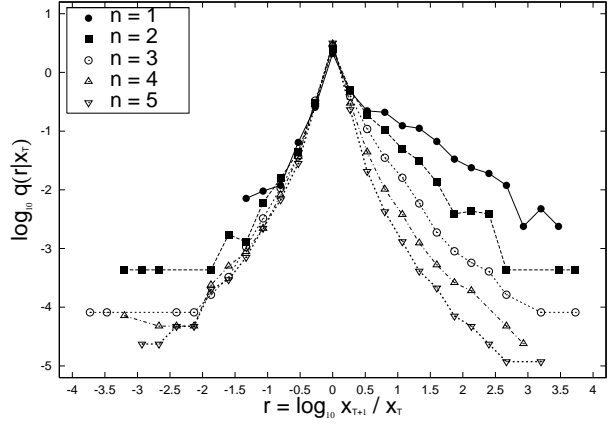}
\caption{\label{Sales-total-GrowthRate-1} Conditional PDFs of sales growth rate in the small- and mid-scale ranges
($10^{3} \le x_T < 10^{5}$).
Here, $x_T$ and $x_{T+1}$ are sales in consecutive years, in thousand yen.}
\end{center}
\end{figure}

\begin{figure}[h]
\begin{center}
\includegraphics[height=6cm]{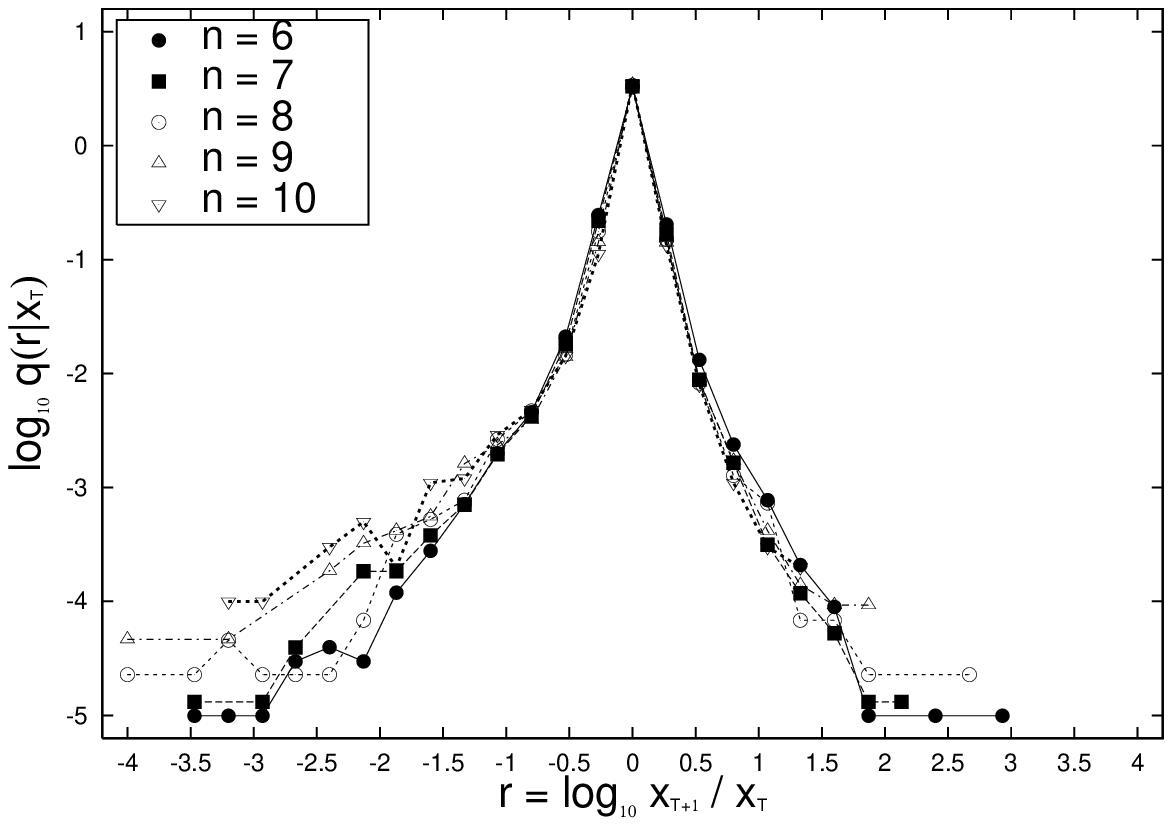}
\caption{\label{Sales-total-GrowthRate-2} Conditional PDFs of sales growth rate in the mid- and large-scale ranges ($10^{5} \le x_T < 10^{7}$).
Here, $x_T$ and $x_{T+1}$ are sales in consecutive years, in thousand yen.}
\end{center}
\end{figure}

\begin{figure}[!h]
\begin{center}
\includegraphics[height=6cm]{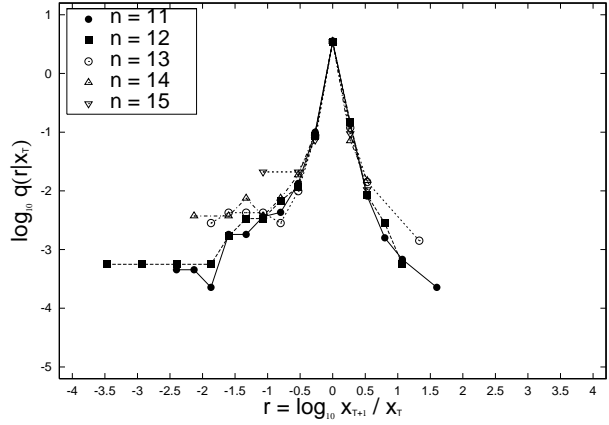}
\caption{\label{Sales-total-GrowthRate-3} Conditional PDFs of sales growth rate in the large-scale range
($10^{7} \le x_T < 10^{9}$).
Here, $x_T$ and $x_{T+1}$ are sales in consecutive years, in thousand yen.}
\end{center}
\end{figure}

Second, we divide the range of the initial value $x_T$ into logarithmically equal bins as
$x_T \in [10^{3+0.4(n-1)},10^{3+0.4n})$ $(n=1,2,\cdots,15)$.
The conditional growth-rate distributions $q(r|x_T)$ are shown in 
Figs.~\ref{Sales-total-GrowthRate-1}--\ref{Sales-total-GrowthRate-3}.
Each growth-rate distribution in 
Figs.~\ref{Sales-total-GrowthRate-1}--\ref{Sales-total-GrowthRate-3}
has curvatures.
It is difficult to approximate the growth-rate 
distributions by the linear approximation
(\ref{approximation1})--(\ref{approximation2}) as in the profits case.
As the simplest extension,
we have added a second-order term with respect to $r$ to express the curvatures
as follows:
\begin{eqnarray}
    \log_{10}q(r|x_T)&=&c(x_T) - t_{+}(x_T)~r+\ln10~u_{+} (x_T)~r^2
    ~~~~~{\rm for}~~r_c > r > 0~,
    \label{approximation3}
    \\
    \log_{10}q(r|x_T)&=&c(x_T) + t_{-}(x_T)~r+\ln10~u_- (x_T)~r^2
    ~~~~~{\rm for}~~- r_c < r < 0~.
    \label{approximation4}
\end{eqnarray}
Note that we must introduce a cut $r_c$ in order to normalize the probability integration
as $\int^{10^{r_c}}_{10^{-r_c}} dR~Q(R|x_T) = 1$,
since Eqs.~(\ref{approximation3}) and (\ref{approximation4})
are quadratic with respect to $r$.  
From this normalization condition, $c(x_T)$ can be expressed by using
$t_{\pm}(x_T)$, $u_{\pm}(x_T)$, and $r_c$.
The expression is quite complicated, and
it is later observed that $c(x_T)$ only slightly depends on $x_T$ in the empirical data.
Therefore, we do not describe the expression here. 

The approximation (\ref{approximation3})--(\ref{approximation4}) is rewritten as
\begin{eqnarray}
    Q(R|x_T)&=&d(x_T)~R^{- 1 - t_{+}(x_T) + u_{+}(x_T) \ln R}~~~~~{\rm for}~~R > 1~,
    \label{SecondNon-Gibrat'sLaw1}
    \\
    Q(R|x_T)&=&d(x_T)~R^{- 1 + t_{-}(x_T) + u_{-}(x_T) \ln R}~~~~~{\rm for}~~R < 1~.
    \label{SecondNon-Gibrat'sLaw2}
\end{eqnarray}
By using this shape, in the case of $R>1$,
detailed balance (\ref{DetailedBalance2}) is reduced to
\begin{eqnarray}
    \frac{\tilde{P}(x_T)}{\tilde{P}(x_{T+1})} = R^{~1+t_{+}(x_T) - t_{-}(x_{T+1})-\left[u_+(x_T)-u_-(x_{T+1})\right]\ln R}~.
    \label{start}    
\end{eqnarray}
By expanding Eq.~(\ref{start}) around $R=1$ with $x_T \to x$ and $x_{T+1} \to R~x$, 
the following five differential equations are obtained:
\begin{eqnarray}
    \Bigl[1+t_{+}(x)-t_{-}(x) \Bigr] \tilde{P}(x) 
        + x~ {\tilde{P}}^{'}(x) = 0~,
    \label{DE3}
\end{eqnarray}
\begin{eqnarray}
    x \left[ {t_{+}}^{'}(x)+{t_{-}}^{'}(x) \right] + 2 \left[ u_{+}(x) - u_{-}(x) \right]=0~,
    \label{DE4}\\
    2~{t_{+}}^{'}(x)+{t_{-}}^{'}(x)+6{u_{+}}^{'}(x)+x \left[2~{t_{+}}^{''}(x)+{t_{-}}^{''}(x) \right]=0~, 
    \label{DE5}\\
    {t_{+}}^{'}(x)+{t_{-}}^{'}(x)+3x \left[{t_{+}}^{''}(x)+{t_{-}}^{''}(x) \right]
    +x^2 \left[{t_{+}}^{(3)}(x)+{t_{-}}^{(3)}(x) \right]=0~,
    \label{DE6}\\
    {t_{+}}^{'}(x)+7x~{t_{+}}^{''}(x)+6x^2~{t_{+}}^{(3)}(x)+x^3~{t_{+}}^{(4)}(x)=0~.
    \label{DE7}
\end{eqnarray}
The same differential equations are obtained for $R < 1$.
Equations~(\ref{DE4})--(\ref{DE7}) uniquely fix the change 
in the growth-rate distribution $t_{\pm}(x)$, $u_{\pm}(x)$ 
as follows:
\begin{eqnarray}
    t_+(x) &=& \frac{\gamma}{3} \ln^3 x + \frac{\beta}{2} \ln^2 x
                + \alpha \ln x + C_{1}~,
    \label{t+}\\
    t_-(x) &=& - \frac{\gamma}{3} \ln^3 x + \frac{\delta - \beta }{2} \ln^2 x
                + (\eta - \alpha) \ln x 
                +C_{2}~,
    \label{t-}\\
    u_+(x) &=& - \frac{\gamma}{6} \ln^2 x - \frac{\delta + \beta}{6} \ln x + C_3~,
    \label{u+}\\
    u_-(x) &=& - \frac{\gamma}{6} \ln^2 x + \frac{2 \delta - \beta}{6} \ln x 
                + C_3 + \frac{\eta}{2}~.
    \label{u-}  
\end{eqnarray}
Now, let us confirm these solutions with the empirical data.

\begin{figure}[b]
\begin{center}
\includegraphics[height=6cm]{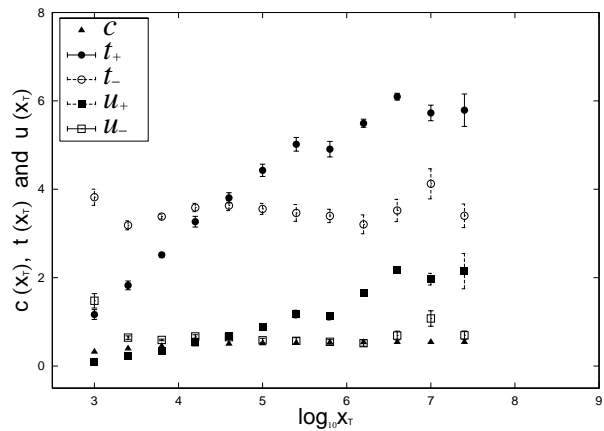}
\caption{\label{Sales-total-Evaluation} Estimations of $c(x_T)$, $t_{\pm}(x_T)$, and $u_{\pm}(x_T)$.
Here, $x_T$ is the lower bound of each bin, in thousand yen, and
$c(x_T)$ is the original value of the growth-rate distribution.
From left, each point on the graph represents $n=1,2, \cdots, 12$.}
\end{center}
\end{figure}

Figure~\ref{Sales-total-Evaluation} shows $t_{\pm}(x_T)$, $u_{\pm}(x_T)$ and $c(x_T)$
estimated by fitting the approximation (\ref{approximation3})--(\ref{approximation4}) 
to each growth-rate distribution in 
Figs.~\ref{Sales-total-GrowthRate-1}--\ref{Sales-total-GrowthRate-3}.
In Fig.~\ref{Sales-total-Evaluation}, $c(x_T)$ is fixed as the empirical value
and $t_{\pm}(x_T)$ and $u_{\pm}(x_T)$ are estimated by using the least-squares method.
For $n=13, 14, 15$ in Fig.~\ref{Sales-total-GrowthRate-3}, there are not sufficient data
points to estimate $t_{\pm}(x_T)$, $u_{\pm}(x_T)$ for $n=14, 15$ or 
to estimate the error bar for $n=13$.
Therefore, data points for $n=13, 14, 15$ are not plotted in 
Fig.~\ref{Sales-total-Evaluation}.

\begin{figure}[!h]
\begin{center}
\includegraphics[height=6cm]{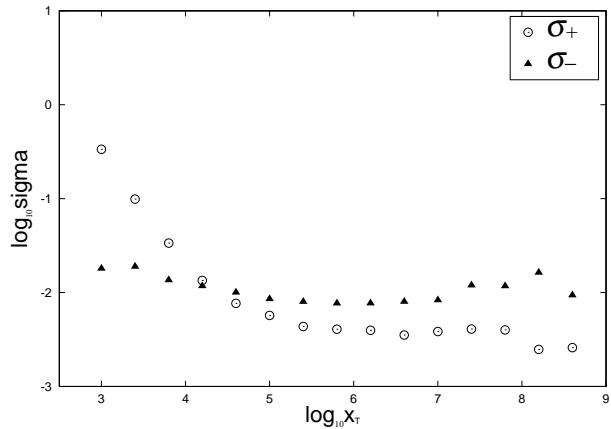}
\caption{\label{var} Estimations of $\sigma_{\pm}(x_T)$.
Here, $x_T$ is the lower bound of each bin, in thousand yen.
From left, each point on the graph represents $n=1,2, \cdots, 15$.}
\end{center}
\end{figure}

On the one hand, for $n=9, 10, \cdots, 15$ $(x_T \ge 10^{6.2})$
in Figs.~\ref{Sales-total-GrowthRate-2} and \ref{Sales-total-GrowthRate-3}, 
the growth-rate distribution hardly changes as $n$ increases.
This means that Gibrat's Law (\ref{Gibrat}) is verified by the empirical data.
We regard $x_T \ge 10^{6.2}$ as the large-scale range
and set $\gamma = \beta = \delta = \alpha = \eta = 0$ in this range
because $t_{\pm}(x_T)$ and $u_{\pm}(x_T)$ do not depend on $x_T$. 
In Fig.~\ref{Sales-total-Evaluation}, the values of $t_{\pm}(x_T)$ and $u_{\pm}(x_T)$ vary 
in this range
because the number of data in Fig.~\ref{Sales-total-GrowthRate-3} 
is statistically insufficient to estimate them
by the least-squares method.
However, by measuring positive and negative standard deviations $\sigma_{\pm}$ of
each growth-rate distribution 
in Figs.~\ref{Sales-total-GrowthRate-1}--\ref{Sales-total-GrowthRate-3},
we confirmed that the growth-rate distribution hardly changes in the large-scale range 
$x_T \ge 10^{6.2}$ (Fig.~\ref{var}). 

On the other hand, in Fig.~\ref{Sales-total-GrowthRate-1},
while the negative growth-rate distribution hardly changes as $n$ increases,
the positive growth-rate distribution gradually decreases.
This is Non-Gibrat's Property in the mid-scale range of sales data.
We should estimate parameters $\gamma, \beta, \delta, \alpha$ and $\eta$
by applying the change in the growth-rate distribution 
(\ref{t+})--(\ref{u-}) to Fig.~\ref{Sales-total-Evaluation}.
However, there are insufficient data points in Fig.~\ref{Sales-total-Evaluation}
for using the least-squares method by polynomial functions (\ref{t+})--(\ref{u-}).
Consequently, as a first-order approximation, we assume that
the negative growth-rate distribution does not depend on $x_T$, 
even in the mid-scale range.
This approximation is guaranteed by Fig.~\ref{var}
because the negative standard deviation $\sigma_-$ hardly changes
compared with the positive standard deviation $\sigma_+$.

In this approximation, the parameters are simplified as 
\begin{eqnarray}
    \gamma=\delta=\beta=0~~~~~ {\rm and}~~~~~ \eta=\alpha~.
    \label{approximation99}
\end{eqnarray}
Only the change in the positive growth-rate distribution $t_+(x_T)$ depends on $x_T$
as follows:
\begin{eqnarray}
    t_+(x)&=&\alpha \ln x + C_{1}~,
    \label{t+2}\\
    t_-(x)&=&C_2~,~~~
    u_+(x)=C_3~,~~~
    u_-(x)=C_3 + \frac{\alpha}{2}~.
    \label{u-2}
\end{eqnarray}
We call this Non-Gibrat's Second Property.

Applying $t_+(x_T)$ (\ref{t+2}) 
to $n=3, 4, 5, 6$ $(10^{3.8} \le x_T < 10^{5.4})$
in Fig.~\ref{Sales-total-Evaluation},
we obtain the rate-of-change parameter $\alpha = 0.68 \pm 0.03$ by 
the least-squares method.
We regard $10^{3.8} \le x_T < 10^{5.4}$ as the mid-scale range of sales.
In this range, $t_{-}(x_T)$ and $u_{\pm}(x_T)$ hardly change compared with
$t_+(x_T)$, so the approximation (\ref{u-2}) is considered relevant.
Nevertheless, the value $\alpha$ estimated by the difference between $u_{+}(x_T)$
and $u_{-}(x_T)$
disagrees with the value estimated by the change in $t_+(x_T)$.
Most likely, this comes from a limitation of the second-order approximation with respect to $r$
(\ref{approximation3})--(\ref{approximation4}). 
To fix this discrepancy,
we may add a third-order term with respect to $r$.
We will consider this point in the conclusion.
In addition, we should note that
the intercept $c(x_T)$ only slightly depends on $x_T$ in the mid- and large-scale ranges
$x_T \ge 10^{3.8}$, as in the profits case.

Using $t_{\pm}(x)$ (\ref{t+})--(\ref{t-}), 
Eq.~(\ref{DE3}) uniquely determines the PDF of $x$ as
\begin{eqnarray}
    P(x) \propto  x^{-(\mu+1)}~
    \exp \Bigl[
                - \frac{\gamma}{6}\ln^4 x
                + \frac{\delta - 2 \beta}{6}\ln^3 x
                -(\alpha-\frac{\eta}{2})\ln^2 x
         \Bigr]~.
    \label{P}    
\end{eqnarray}
Here, we regard $d(x)$ in $\tilde{P}(x)=d(x)~P(x)$ as a constant and 
denote $\mu=C_1 - C_2$.
The solutions (\ref{t+})--(\ref{u-}) and (\ref{P}) 
satisfy Eq.~(\ref{start}) beyond perturbation around $R = 1$, so
these are not only necessary but also sufficient.
In the approximation (\ref{approximation99}), the PDF is reduced to
\begin{eqnarray}
    P(x)\propto x^{-(\mu+1)}~\exp \left[-\frac{\alpha}{2} \ln^2 x \right]~.
    \label{P2}
\end{eqnarray}

\begin{figure}[t]
\begin{center}
\includegraphics[height=6cm]{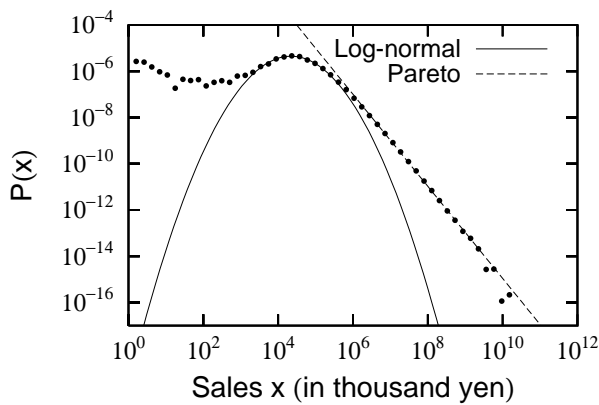}
\caption{\label{SalesDistribution-total} 
A PDF of sales in the database.
Pareto's Law is observed in the large-scale range ($x > 10^{6.2}$)
and the log-normal distribution in the mid-scale range ($10^{3.8} \le x < 10^{5.4}$).}
\end{center}
\end{figure}

Figure~\ref{SalesDistribution-total} shows that the resulting PDF (\ref{P2})
fits correctly with the empirical sales data.
In the large-scale range ($\alpha=0$), 
the PDF (\ref{P2}) behaves as Pareto's Law (\ref{Pareto}).
The Pareto index is estimated as approximately $\mu \sim 1$
in the large-scale range ($x \ge 10^{6.2}$) of Fig.~\ref{SalesDistribution-total}.
In the mid-scale range, the PDF (\ref{P2}) behaves as the 
log-normal distribution (\ref{Log-normal}) in the same manner as in the profits case.
Applying the PDF (\ref{P2}) to the mid-scale range ($10^{3.8} \le x < 10^{5.4}$)
of Fig.~\ref{SalesDistribution-total}, we obtain the rate-of-change parameter
$\alpha = 0.65 \pm 0.04$
by using the least-squares method.
This is consistent with 
the value estimated by the change in $t_{+}(x_T)$ ($\alpha = 0.68 \pm 0.03$).
From these results, we conclude that Non-Gibrat's Second Property is also confirmed 
by the empirical data.

\section{Conclusion}
In this study, we have employed exhaustive business data on Japanese firms 
that nearly cover not only the entire large-scale range but also the entire mid-scale range in terms of firm size.
Using this newly assembled database,
we first reconfirmed the previous analyses for profits data \cite{Ishikawa} as described below.
In the mid-scale range, the log-normal distribution is derived from detailed balance and 
from Non-Gibrat's First Property.
In Non-Gibrat's First Property, the probability of positive growth decreases and the  probability of negative growth increases symmetrically as the initial value $x_T$ increases.
Under detailed balance, this change is uniquely reduced from the shape of the growth-rate distribution, which is linear in log-log scale.

Second, the following findings were reported with respect to sales data.
Detailed balance is also observed in the mid- and large-scale ranges of sales data.
The growth-rate distribution of sales has wider tails than the linear growth-rate distribution of profits in log-log scale.
In the mid-scale range, while the probability of negative growth hardly changes as the initial value $x_T$ increases, the probability of positive growth gradually decreases. This feature is different from Non-Gibrat's First Property observed in the profits data.
We have approximated the growth-rate distribution with curvatures by a quadratic function. In addition, from an empirical observation, we have imposed the condition that the negative growth-rate distribution does not depend on $x_T$, even in the mid-scale range. 
Under detailed balance, these approximations and conditions uniquely lead to a decrease in positive growth. We call this Non-Gibrat's Second Property.
In the mid-scale range, the log-normal distribution is also derived from detailed balance and from Non-Gibrat's Second Property. These results are confirmed by the empirical data.

In this study, it was clarified that
the shape of the growth-rate distribution of sales is different from that of profits.
It was also demonstrated that this difference is closely related to 
the difference between two kinds of Non-Gibrat's Properties in the mid-scale range.
The growth-rate 
distribution of income of firms is approximated by a linear function in
log-log scale as in the profits data. 
The growth-rate distributions of assets, the number of employees, and personal income 
have wider tails than a linear function in log-log scale, as in the sales data.
If we obtained exhaustive data that include the mid-scale range,
Non-Gibrat's First Property would probably be observed in the income data of firms,
while Non-Gibrat's Second Property would probably be observed in the assets, 
the number of employees, and the personal income data.

We have not determined what makes the difference between the shapes of the growth-rate distributions.
However, this difference is probably related to the following factors \cite{Economics}.
Income and profits of firms are calculated by a subtraction
of total expenditures from total sales in a rough estimate. 
Assets and sales of firms, the number of
employees, and personal income are not calculated by any subtraction.

Let us consider the distribution of added values, the sum of which is GDP.
Clearly, added values are calculated by some subtraction.
If we obtained exhaustive data of added values,
Non-Gibrat's First Property would certainly be observed.
It has been reported that the growth-rate distribution of GDPs of countries
is linear in log-log scale (for instance \cite{Canning}).
This report reinforces that speculation. 
The results in this paper should be carefully considered in cases 
where governments and firms discuss strategies of growth.

Finally, we consider a method to fix the inconsistency by which the rate-of-change parameter
$\alpha$ is not estimated by the difference between $u_{\pm}(x_T)$ (\ref{u-2}).
Let us add not only the second-order term with respect to $r$ but also a third-order
term as follows:
\begin{eqnarray}
    \log_{10}q(r|x_T)&=&c(x_T) - t_{+}(x_T)~r+\ln10~u_{+} (x_T)~r^2 -\ln^2 10~v_{+}(x_T)~r^3
    ~~~~~{\rm for}~~r > 0~,
    \label{approximation5}
    \\
    \log_{10}q(r|x_T)&=&c(x_T) + t_{-}(x_T)~r+\ln10~u_- (x_T)~r^2 +\ln^2 10~v_{-}(x_T)~r^3
    ~~~~~{\rm for}~~r < 0~.
    \label{approximation6}
\end{eqnarray}
In the same manner as in the previous section, under detailed balance,
coefficients $t_{\pm}(x)$, $u_{\pm}(x)$,
and $v_{\pm}(x)$ are uniquely obtained as follows:
\begin{eqnarray}
    t_+(x) &=& \frac{\zeta}{5} \ln^5 x + \frac{\epsilon}{4} \ln^4 x 
                + \frac{\gamma}{3} \ln^3 x + \frac{\beta}{2} \ln^2 x
                + \alpha \ln x + C_{1}~,
    \label{t+3}\\
    t_-(x) &=&-\frac{\zeta}{5} \ln^5 x + \frac{\kappa - \epsilon}{4} \ln^4 x  
                + \frac{\theta - \gamma}{3} \ln^3 x + \frac{\delta - \beta}{2} \ln^2 x
                + (\eta - \alpha) \ln x 
                +C_{2}~,
    \label{t-3}\\
    u_+(x) &=&-\frac{\zeta}{5} \ln^4 x  
                - \frac{4\epsilon + 3\kappa}{20} \ln^3 x 
                - (\lambda + \frac{2\gamma + \theta}{12}) \ln^2 x 
                + (\nu - \frac{\delta + \beta}{6}) \ln x + C_3~,
    \label{u+3}\\
    u_-(x) &=&-\frac{\zeta}{5} \ln^4 x  
                - \frac{4\epsilon - 7\kappa}{20} \ln^3 x  
                - (\lambda + \frac{2\gamma - 5\theta}{12}) \ln^2 x 
                + (\nu + \frac{2\delta - \beta}{6}) \ln x   \nonumber\\
                &&+ C_3+ \frac{\eta}{2}~,
    \label{u-3}\\
    v_+(x) &=& \frac{\zeta}{15} \ln^3 x + \frac{2\kappa+\epsilon}{20} \ln^2 x + \lambda \ln x+C_4~,
    \label{v+3}\\
    v_-(x) &=&-\frac{\zeta}{15} \ln^3 x + \frac{3\kappa-\epsilon}{20} \ln^2 x 
                - (\lambda - \frac{\theta}{6})\ln x+C_4+\mu~.
    \label{v-3}      
\end{eqnarray}

By imposing the condition that the negative growth-rate distribution does not depend on $x_T$
even in the mid-scale range, these are simplified as follows:
\begin{eqnarray}
    t_+(x) &=& \frac{\beta}{2} \ln^2 x
                + \alpha \ln x + C_{1}~,
    ~~~~~
    t_-(x) = C_{2}~,
    \label{t4}\\
    u_+(x) &=&-\frac{\beta}{2} \ln x + C_3~,
    ~~~~~
    u_-(x) = C_3+ \frac{\alpha}{2}~,
    \label{u4}\\
    v_+(x) &=& C_4~,
    ~~~~~
    v_-(x) = C_4-\frac{\beta}{6}~.
    \label{v4}      
\end{eqnarray}
The results in the previous sections (\ref{t+2}) and (\ref{u-2}) correspond to
a special case $\beta=0$, $C_4=0$ in Eqs.~(\ref{t4})--(\ref{v4}).
In the previous section, it was difficult to estimate $\alpha$ by the difference in
$u_{\pm}(x)$. In the expressions (\ref{t4})--(\ref{v4}), 
this discrepancy is probably solved with a negative $\beta$.
Note that Eqs.~(\ref{t+})--(\ref{u-}) cannot be reduced to Eqs.~(\ref{t4}) and (\ref{u4}) 
in any parameterization.

It is technically difficult to estimate $t_{\pm}(x)$, $u_{\pm}(x)$, and $v_{\pm}(x)$
by approximating the growth-rate distribution
by the cubic function (\ref{approximation5})--(\ref{approximation6}) 
and to estimate $\beta$ and $\alpha$ fitting Eqs.~(\ref{t4})--(\ref{v4})
by the least-squares method.
At the same time, under the approximation by the cubic function 
(\ref{approximation5})--(\ref{approximation6}),
the integration $\int^{\infty}_{0} dR~Q(R|x_T)$ converges without a cut $r_c$,
as in the linear approximation.
Because this work involves difficulties as well as advantages,  
we will investigate the above issues in the near future.


\section*{Acknowledgments}
The authors thank the Research Institute of Economy, Trade and Industry, IAA (RIETI) for supplying the data set used in this work.
This study was produced from the research the authors conducted as
members of the Program for Promoting Social Science Research Aimed at
Solutions of Near-Future Problems, ``Design of Interfirm Networks to
Achieve Sustainable Economic Growth.'' 
This work was
supported in part by a Grant-in-Aid for Scientific Research (C) (No. 20510147) from
the Ministry of Education, Culture, Sports, Science and Technology, Japan.
Takayuki Mizuno was supported by funding from the Kampo Foundation 2009.

\end{document}